\documentclass[graphicx, 12pt]{article}
\usepackage{indentfirst}

\usepackage{graphicx}
\usepackage{caption}
\usepackage{subfigure}

\begin{document}

\small
\hoffset=-1truecm
\voffset=-2truecm
\title{\bf The tunneling radiation of a black hole with a $f(R)$
global monopole under generalized uncertainty principle}

\author{Lingshen Chen\hspace {1cm} Hongbo Cheng\footnote
{E-mail address: hbcheng@ecust.edu.cn}\\
Department of Physics,\\ East China University of Science and
Technology,\\ Shanghai 200237, China\\}

\date{}
\maketitle

\begin{abstract}
The Parikh-Kraus-Wilczeck tunneling radiation of black hole
involving a $f(R)$ global monopole is considered based on the
generalized uncertainty principle. The influences from global
monopole, $f(R)$ gravity and the corrections to the uncertainty
appear in the expression of black hole entropy difference. It is
found that the global monopole and the revision of general
relativity both hinder the black hole from emitting the photons.
The two parts as corrections to the uncertainty make the entropy
difference of this kind of black hole larger or smaller
respectively.
\end{abstract}

\vspace{7cm} \hspace{1cm} PACS number(s): 04.70.Dy, 03.65.Sq, 04.62.+v

\noindent \textbf{I.\hspace{0.4cm}Introduction}

The black holes were thought as perfect absorbers classically
without emitting anything [1,2]. More than forty years ago, S. W.
Hawking put forward that a black hole can radiate particles
quantum-mechanically and the radiation spectrum is purely thermal,
while several derivations of this kind of emission emerge [3-5].
Afterwards the Hawking radiation was perceived as a semi-classical
tunneling process, leading an alternative method [6-9]. This
tunneling formalism due to the imaginary part of action for
classically forbidden region of emission across the horizon
attracted a lot of attention [10-16]. The discussion on Hawking
radiation based on the semi-classical tunneling proposed by Kraus
et. al. has been used to explore many kinds of black hole
spacetimes such as BTZ black hole [17-20], a series of Taub-NUT
black holes [21], Kerr-Newman black holes [22-24], Godel black
hole [25], etc..

In the development of quantum gravity some new proposals beyond
the previous framework generated. The quantum gravity needs a
minimal length of the order of the Planck length [26-32]. Further
the generalized uncertainty principle (GUP) as generalization of
Heisenberg's scheme was initiated to modify the quantum mechanics
[33]. The following efforts were contributed to the GUP because of
the existence of a minimum measurable length scale [27, 34-39].

There has been much interest in the quantum gravitational
influence within the frame of GUP. In particular, the relation
between the entropy of black hole and a minimal length as quantum
gravity scale was derived and estimated [40, 41]. The effects from
the GUP on the corrected Beckenstein-Hawking black hole entropy in
the higher dimensional spacetime was investigated while the black
hole radiation was discussed with the help of the tunneling
formalism [40]. The Hawking tunneling radiation from black holes
involving GUP corrections was also scrutinized in the world with
extra dimensions [41].

During the vacuum phase transition in the early universe, several
types of topological defects such as domain walls, cosmic strings
and monopoles may generate due to a breakdown of local or global
gauge symmetries [42, 43]. As a spherically symmetric topological
object, a global monopole appeared in the phase transition of a
system involving a self-coupling triplet of a scalar field whose
original global $O(3)$ symmetry is spontaneously broken to $U(1)$
[42, 43]. The metric outside a massive source with a global
monopole was investigated and the distinct properties were shown
that the surroundings has a solid angle leading all light rays
deflected at the same angle although the monopole exerts
practically no gravitational force on nonrelativistic matter [44].
The theory of $f(R)$ gravity for the acceleration of the universe
modifies the description of spacetime significantly [45]. This
kind of theory is utilized to explain the accelerated-inflation
problem without dark matter or dark energy [46-48]. T. R. P.
Carames et al discussed the gravitational field of massive source
swallowing a global monopole within the frame of $f(R)$ gravity
theory to put forward a parameter $\psi_{0}$ associated with the
corrections from the gravity. It is interesting that this
nonvanishing modified parameter also forms an outer horizon as a
boundary of the universe subject to the $f(R)$ monopole metric
[49, 50]. The thermodynamic quantities of the $f(R)$ monopole
black hole were estimated [51]. Recently, F. B. Lustosa et al
generalize the metric where $\frac{df(R)}{dR}=F(R)$ is a power law
function of the radial coordinate of the global monopole
spacetime. They also analyze the thermodynamics of this kind of
black holes [52]. The strong gravitational lensing for a massive
source with a $f(R)$ global monmopole was discussed analytically
[53].

It is significant to study the radiation of the black holes with
$f(R)$ global monopole based on the GUP in addition to their
gravitational lensing and thermodynamics. In the case of minimal
quantum gravity order, black holes have to be explored while the
Heisenberg's uncertainty is generalized. We relate the revision
from quantum gravitation with the deviation of general relativity
in the presentation of the entropy difference of this kind of
black holes by means of the tunneling formalism. We derive and
calculate the tunneling probability consisting of the black hole
entropy having something to do with the factors mentioned just
now. Our discussions and conclusions are listed in the end.

\vspace{0.8cm} \noindent \textbf{II.\hspace{0.4cm}The entropy
difference of a  radiating black hole with a $f(R)$ global
monopole}

We are going to investigate the modifications from several
directions such as global monopole, $f(R)$ scheme and GUP. Now we
start to focus on the entropy of black hole with global monopole
in the $f(R)$ theory. The spherical symmetric line element for the
gravitational field with global monopole in the $f(R)$ theory is
adapted as follow [49, 50, 52],

\begin{equation}
ds^{2}=A(r)dt^{2}-B(r)dr^{2}-r^2(d\theta^{2}+\sin^{2}\theta
d\varphi^{2})
\end{equation}

\noindent where

\begin{equation}
A(r)=B^{-1}(r)=1-8\pi G\eta^{2}-\frac{2GM}{r}-\psi_{0}r
\end{equation}

\noindent Here $G$ is the Newton constant. As a monopole
parameter, $\eta$ is of the order $10^{16}GeV$ in a typical grand
unified theory, leading $8\pi G\eta^{2}\approx10^{-5}$ [43, 44].
$M$ is mass parameter. The factor $\psi_{0}$ reflects the
extension of the standard general relativity. It is obvious that
the term $\psi_{0}r$ in the metric above is linear, which is
certainly different from the structures of de Sitter spacetime and
the Reissner-Nordstrom metric, etc. [49, 50]. As roots of $A(r)=0$
for metric (1), there exists an inner radius,

\begin{equation}
r_{-}=\frac{1-8\pi G\eta^{2}-\sqrt{(1-8\pi
G\eta^{2})^{2}-8GM\psi_{0}}}{\psi_{0}}
\end{equation}

\noindent and an outer ones,

\begin{equation}
r_{+}=\frac{1-8\pi G\eta^{2}+\sqrt{(1-8\pi
G\eta^{2})^{2}-8GM\psi_{0}}}{\psi_{0}}
\end{equation}

\noindent If the modified parameter $\psi_{0}$ vanishes, the outer
horizon will disappear. According to Ref. [3-5], the Hawking
temperature for the black hole from Eq. (1) and Eq. (2) is a
function of variables like $\eta$, $\psi_{0}$ etc.,

\begin{equation}
T_{H}=\frac{1}{4\pi}(\frac{1-8\pi G\eta^{2}}{r_{H}}-2\psi_{0})
\end{equation}

\noindent The Beckenstein-Hawking entropy may be derived from the
Hawking temperature in virtue of the following thermodynamics
relation [3-5, 9],

\begin{equation}
T_{H}=\frac{dE}{dS}\approx\frac{dM}{dS}
\end{equation}

\noindent The difference between the initial and the final
magnitudes of the entropy of the $f(R)$ black hole containing the
global monopole in the emission process can be approximated as,

\begin{equation}
\bigtriangleup S\approx-\frac{4\pi G}{(1-8\pi G\eta^{2})^{2}}
[M^{2}-(M-\hbar\omega)^{2}]-\frac{16\pi G^{2}\psi_{0}}{(1-8\pi
G\eta^{2})^{4}} [M^{3}-(M-\hbar\omega)^{3}]
\end{equation}

\noindent where $\omega$ is a shell of energy moving along the
geodesics in the spacetime with metric (1) [9]. The higher order
of typical grand unified theory or farther away from standard
general relativity will lead larger absolute value of negative
entropy difference. According to Ref. [9], the relation
probability of the black hole can be demonstrated as,

\begin{equation}
\Gamma\sim e^{\bigtriangleup S}
\end{equation}

\noindent The existence of global monopole in the black hole or
the deviation from general relativity damps the emission of the
black hole.

\vspace{0.8cm} \noindent \textbf{III.\hspace{0.4cm}The entropy
difference of a  radiating black hole with a $f(R)$ global
monopole under generalized uncertainty principle}

Here we turn our discussions in the context of GUP. The
Heisenberg's uncertainty principle is generalized within the
microphysics regime as [36, 39, 40, 54-64],

\begin{equation}
\bigtriangleup x\bigtriangleup p\geq\frac{\hbar}{2}[1-\frac{\alpha
l_{p}}{\hbar}\bigtriangleup p+(\frac{\beta
l_{p}}{\hbar})^{2}\bigtriangleup p^{2}]
\end{equation}

\noindent leading

\begin{equation}
y_{-}\leq y\leq y_{+}
\end{equation}

\noindent where

\begin{eqnarray}
y_{\pm}=(\frac{l_{p}}{\hbar}\bigtriangleup p)_{\pm}\hspace{6.5cm}\nonumber\\
=\frac{1}{2\beta^{2}}(\alpha+\frac{2\bigtriangleup x}{l_{p}})
\pm\frac{1}{2\beta^{2}}(\alpha+\frac{2\bigtriangleup
x}{l_{p}})\sqrt{1-(\frac{2\beta}{\alpha+\frac{2\bigtriangleup
x}{l_{p}}})^{2}}
\end{eqnarray}

\noindent where $\alpha$ and $\beta$ are dimensionless positive
parameters, Planck length shown as $l_{p}=\sqrt{\frac{\hbar
G}{c^{3}}}$ while the velocity of light $c$. The terms with the
Newtonian constant $G$ provide the uncertainty with the
gravitational effects. According to the procedure of Ref. [40, 41,
65, 66], we introduce,

\begin{equation}
\bigtriangleup p'=\frac{\hbar}{l_{p}}y_{-}
\end{equation}

\noindent Approximately the uncertainty in the momentum is [65],

\begin{equation}
\bigtriangleup p'\approx\frac{\hbar}{\alpha l_{p}+2\bigtriangleup
x}
\end{equation}

\noindent We substitute the approximation (13) into the GUP (9) to
estimate the distance interval [65],

\begin{equation}
\bigtriangleup x'\approx\bigtriangleup x[1+\frac{(\beta
l_{p})^{2}}{2\bigtriangleup x(\alpha l_{p}+2\bigtriangleup x)}]
\end{equation}

\noindent The original size of black hole is chosen as the lower
bound on the region like $\bigtriangleup x=2r_{H}$ [40, 41], so
the Hawking temperature (5) can be changed as follow,

\begin{equation}
T_{H}=\frac{1}{2\pi}(\frac{1-8\pi G\eta^{2}}{\bigtriangleup
x}-\psi_{0})
\end{equation}

\noindent If the black hole exists without global monopole and the
$f(R)$ corrections, the temperature will return to that of
Schwarzschild case [40, 41]. Like Ref. [65], we choose the
distance interval in the temperature (15) as $\bigtriangleup x'$
expressed in Eq. (14) to obtain,

\begin{eqnarray}
T'_{H}=\frac{1}{2\pi}(\frac{1-8\pi G\eta^{2}}{\bigtriangleup
x'}-\psi_{0})\nonumber\\
\approx T_{H}-\frac{(1-8\pi G\eta^{2})(\beta
l_{p})^{2}}{4\pi\bigtriangleup x^{2}(\alpha l_{p}+2\bigtriangleup
x)}
\end{eqnarray}

\noindent The corrections from GUP make the Hawking temperature
lower. We make use of the thermodynamic relation (6) [3-5, 9] to
obtain the corrected entropy difference of the radiating black
hole with $f(R)$ global monopole as,

\begin{eqnarray}
\bigtriangleup S'=\bigtriangleup S'(\eta, \psi_{0}, \alpha,
\beta)\hspace{10.5cm}\nonumber\\
=\frac{\pi}{G}(r'^{2}_{H}-r_{H}^{2})\hspace{12cm}\nonumber\\
-\frac{\pi}{G}\frac{(1-8\pi
G\eta^{2})(\beta l_{p})^{2}}{16\psi_{0}r_{H}^{2}+4[\psi_{0}\alpha
l_{p}-2(1-8\pi G\eta^{2})]r_{H}+[\psi_{0}(\beta
l_{p})^{2}-2(1-8\pi G\eta^{2})\alpha l_{p}]}(r'_{H}-r_{H})
\end{eqnarray}

\noindent where we replace the mass parameter $M$ in the horizon
(3) as $M-\hbar\omega$ to obtain the horizon at the end of the
process that black hole emits a photon [9]. It can be checked that
the corrected entropy difference will return to that of Eq. (7)
without generalizing the uncertainty. In order to make clear how
the generalization of Heisenberg Uncertainty affect the tunneling
probability, we compare the entropy difference $\bigtriangleup S'$
with the difference for Schwarzschild black hole like
$\bigtriangleup S_{0}\approx-8\pi GM\hbar\omega$ [9] to obtain,

\begin{eqnarray}
\frac{\bigtriangleup S'}{\bigtriangleup S_{0}}\hspace{11.5cm}\nonumber\\
=\frac{1}{(1-8\pi G\eta^{2})^{2}}+\frac{6GM\psi_{0}}{(1-8\pi
G\eta^{2})^{4}}\hspace{7cm}\nonumber\\
-\frac{(\beta l_{p})^{2}}{16\psi_{0}r_{H}^{2}+4[\psi_{0}\alpha
l_{p}-2(1-8\pi G\eta^{2})]r_{H}+[\psi_{0}(\beta
l_{p})^{2}-2(1-8\pi G\eta^{2})\alpha l_{p}]}\nonumber\\
\times[\frac{1}{4GM}+\frac{\psi_{0}}{(1-8\pi
G\eta^{2})^{2}}+\frac{6GM\psi_{0}^{2}}{(1-8\pi
G\eta^{2})^{4}}]\hspace{3cm}
\end{eqnarray}

\noindent The influence from GUP on the entropy difference in the
case of black hole including $f(R)$ global monopole is shown
graphically. The dependence of the ratio $\frac{\triangle
S'}{\triangle S}$ formulated in Eq. (18) on the variables is
plotted in the figures. There are two factors appeared as $\alpha$
- term and $\beta$ - term respectively in the GUP in Eq. (9). It
is found that the greater correction denoted as $\alpha$ leads the
absolute value of $\bigtriangleup S'$ smaller, which retards the
radiation of the black holes in Fig. 1. The other figure labelled
as Fig. 2 shows that the growth of the other parameter $\beta$
will promote the tunnel process of the black hole because of the
larger absolute value of entropy difference.

\vspace{0.8cm} \noindent \textbf{IV.\hspace{0.4cm}Discussion}

The main results of this paper is Eq. (17) representing the
entropy difference of the black hole containing $f(R)$ global
monopole controlled by the generalized uncertainty principle
during its radiation. The discussions show that the global
monopole inside the black hole as well as $f(R)$ amendments will
decrease the black hole's tunneling probability. The GUP as a
description of the quantum gravitational influence can also be
explored in the process of black hole radiation because its
deviations have something to do with the tunneling probability
[40, 41]. The GUP [36, 39, 40, 54-64] that we employ here consists
of two terms modifying the usual uncertainty and the effects of
the two terms multiplied by $\alpha$ and $\beta$ respectively are
opposite each other. We find that the larger factor $\alpha$
causes the negative term to lower the tunnel radiation of black
hole in the case of black hole with a deficit solid angle outside
in the $f(R)$ scheme. The emission of this kind of black hole is
advanced in favour of the greater parameter $\beta$ as a
coefficient of the positive part. Further works that the GUP is
discussed in other directions proceed.

\vspace{1cm}
\noindent \textbf{Acknowledge}

This work is supported by NSFC No. 10875043.

\newpage
\begin{figure}
\setlength{\belowcaptionskip}{10pt} \centering
\includegraphics[width=15cm]{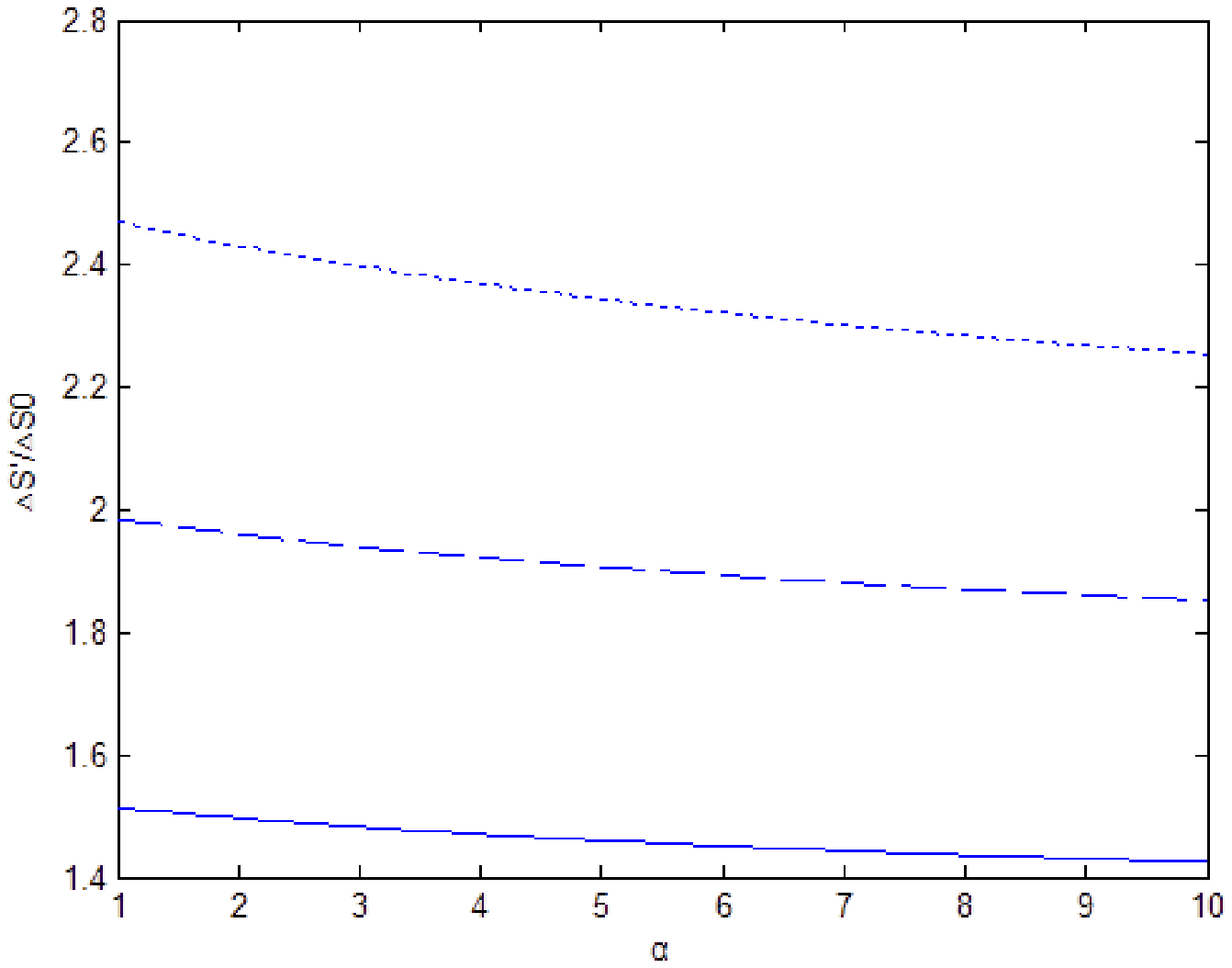}
\caption{The solid, dashed, dot curves of the dependence of the
ratio $\frac{\bigtriangleup S'}{\bigtriangleup S_{0}}$ on $\alpha$
for $\psi_{0}=0.01, 0.05, 0.08$ respectively with $\beta=3.5$ and
for simplicity $8\pi G\eta^{2}=0.1, G=1=M=l_{p}=1$.}
\end{figure}

\newpage
\begin{figure}
\setlength{\belowcaptionskip}{10pt} \centering
\includegraphics[width=15cm]{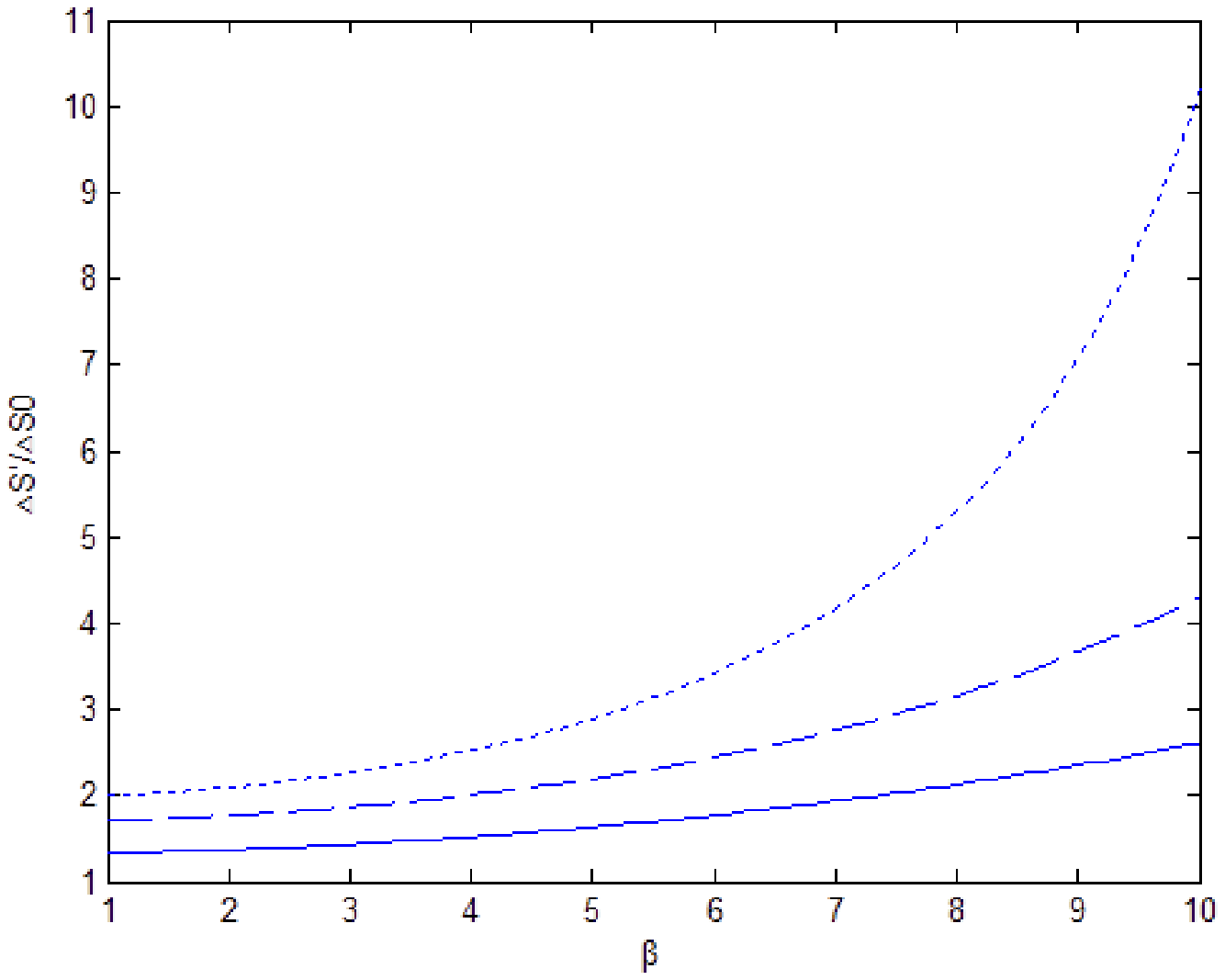}
\caption{The solid, dashed, dot curves of the dependence of the
ratio $\frac{\bigtriangleup S'}{\bigtriangleup S_{0}}$ on $\beta$
for $\psi_{0}=0.01, 0.05, 0.08$ respectively with $\alpha=3.5$ and
for simplicity $8\pi G\eta^{2}=0.1, G=1=M=l_{p}=1$.}
\end{figure}

\end{document}